\documentclass[conference]{IEEEtran}
\IEEEoverridecommandlockouts
\usepackage[T1]{fontenc}
\usepackage{cite}
\usepackage{amsmath,amssymb,amsfonts}
\usepackage{multirow}
\usepackage{graphicx}
\usepackage{subcaption}
\usepackage{textcomp}
\usepackage{xcolor}
\usepackage{tabularray}
\usepackage{todonotes}
\usepackage{booktabs}
\usepackage{comment}
\usepackage{booktabs}
\usepackage{tabularx}
\usepackage[font=footnotesize]{caption}
\usepackage{pgfplots} 
\usetikzlibrary{patterns,arrows,chains,fit,positioning,shapes.symbols,shapes.geometric}
\usepackage{algorithm}
\usepackage{algpseudocode}
\usepackage{sansmath}
\usepackage{authblk}
\usepackage{hyperref}
\usepackage{multirow}
\hypersetup{
	colorlinks=true,
	linkcolor=blue,
	filecolor=magenta,      
	urlcolor=blue,
	citecolor=blue
}

\pgfplotsset{compat=1.17,tick label style = {font=\sansmath\sffamily}}
\def\BibTeX{{\rm B\kern-.05em{\sc i\kern-.025em b}\kern-.08em
T\kern-.1667em\lower.7ex\hbox{E}\kern-.125emX}}
\definecolor{myred}{RGB}{199,101,93}
\newcounter{myframe}
\tikzset{
common/.style={
  text width=4cm,
  align=center,
  text depth=3ex,
  inner ysep=0pt,
  thick},
  mynode1/.style={
  draw=#1,  align=center,
  on chain,
  text width=3.5cm,minimum height = 1.1cm,inner ysep=-10pt, 
  join
  },
mynode/.style={
  common,
  draw=#1,
  on chain,
  text width=3.25cm,minimum height = 0.75cm, 
  join
  },
  mynodelast/.style={
  draw=#1,  align=center,
  on chain,
  text width=1.9cm,minimum height = 0.8cm, 
  join
  },
mytitle/.style={
  common,
  draw=none,
  fill=#1!20},
every join/.style={->,black},
begin/.style={
  signal,  text width=1cm,   align=center,
  draw=green!65!black,
  signal to=east and west},
accept/.style={
   align=center,
  shape aspect=2,inner ysep=-10pt,
  diamond,text width=2.25cm },
end/.style={
  align=center,
  draw,
  rounded corners=12pt,
  text width=1.5cm}
}

\usepackage[pscoord]{eso-pic} 
\newcommand{\placetextbox}[3]{
    \setbox0=\hbox{#3}
    \AddToShipoutPictureFG*{
        \put(\LenToUnit{#1\paperwidth},\LenToUnit{#2\paperheight}){
            \vtop{{\null}\makebox[0pt][c]{#3}}}
        }
}
\placetextbox{.2}{0.055}{978-3-903176-62-1~\copyright~2024 IFIP}

\begin{document}

\setlength{\textfloatsep}{4pt plus 1.0pt minus 2.0pt}
\setlength{\dbltextfloatsep}{4pt plus 1.0pt minus 2.0pt}
\setlength{\skip\footins}{4pt}
\addtolength{\abovecaptionskip}{-0.075in}
\addtolength{\belowcaptionskip}{-0.05in}

\makeatletter
\let\origsection\section
\renewcommand\section{\@ifstar{\starsection}{\nostarsection}}

\newcommand\nostarsection[1]
{\sectionprelude\origsection{#1}\sectionpostlude}

\newcommand\starsection[1]
{\sectionprelude\origsection*{#1}\sectionpostlude}

\newcommand\sectionprelude{%
  \vspace{-0.45em}
}

\newcommand\sectionpostlude{%
  \vspace{-0.35em}
}
\makeatother

\title{Liquid Neural Network-based Adaptive Learning vs. Incremental Learning for Link Load Prediction amid Concept Drift due to Network Failures\\

\thanks{\footnotesize O. Ayoub and D. Andreoletti are co-first authors of this paper. This work was supported by National Science Center, Poland under Grants 2018/31/D/ST6/03041 and 2019/35/B/ST7/04272, the NAWA STER Programme Internationalisation of Wroclaw University of Science and Technology Doctoral School and by the European Union under the Italian National Recovery and Resilience Plan (NRRP) of NextGenerationEU, partnership on “Telecommunications of the Future” (PE00000001 - program “RESTART”).}}

\author{Omran Ayoub$^1$, Davide Andreoletti$^1$, Aleksandra Knapińska$^2$, Róża Goścień$^2$, Piotr Lechowicz$^{2,3}$, \\ Tiziano Leidi$^1$, Silvia Giordano$^1$, Cristina Rottondi$^4$, Krzysztof Walkowiak$^2$}

\affil{\small $^1$University of Applied Sciences of Southern Switzerland, Switzerland $^2$Wroc\l{}aw University of Science and Technology, Poland \\
$^3$Chalmers University of Technology, Sweden $^4$Politecnico di Torino, Italy}

\maketitle

\begin{abstract}
Adapting to concept drift is a challenging task in machine learning, which is usually tackled using incremental learning techniques that periodically re-fit a learning model leveraging newly available data. A primary limitation of these techniques is their reliance on substantial amounts of data for retraining. The necessity of acquiring fresh data introduces temporal delays prior to retraining, potentially rendering the models inaccurate if a sudden concept drift occurs in-between two consecutive retrainings. In communication networks, such issue emerges when performing traffic forecasting following a~failure event: post-failure re-routing may induce a drastic shift in distribution and pattern of traffic data, thus requiring a timely model adaptation. In this work, we address this challenge for the problem of traffic forecasting and propose an approach that exploits adaptive learning algorithms, namely, liquid neural networks, which are capable of self-adaptation to abrupt changes in data patterns without requiring any retraining. Through extensive simulations of failure scenarios, we compare the predictive performance of our proposed approach to that of a reference method based on incremental learning. Experimental results show that our proposed approach outperforms incremental learning-based methods in situations where the shifts in traffic patterns are drastic. %Overall, our study provides valuable indications to network managers on how to confront concept drift issues for traffic forecasting under failure scenario.
\end{abstract}


\section{Introduction}
Network traffic prediction represents a foundational problem in optical network design due to its significant implications for the overall performance and efficiency of the network \cite{ref1, ref2}. Currently, network operators perform traffic prediction and then feed forecast values as inputs to optimization algorithms responsible for fine-tuning the resource allocation and the service provisioning in the network \cite{knapinska2023link}. 

To accurately predict network traffic, operators typically rely on machine learning (\textsc{ml}) algorithms. More specifically, operators train \textsc{ml} models to extrapolate future trends in traffic based on historical data reflecting past traffic patterns \cite{ref1, ref2}. However, the dynamic nature of traffic in networks, coupled with the continuous deployment of emerging services, introduces a plethora of challenges that have garnered widespread research attention. Among these challenges, continuous adaptation of \textsc{ml} models to align with evolving traffic patterns is of particular concern. 

\begin{figure}
\hspace{-1.0cm}
  %\centering
  \includegraphics[width=0.575\textwidth]{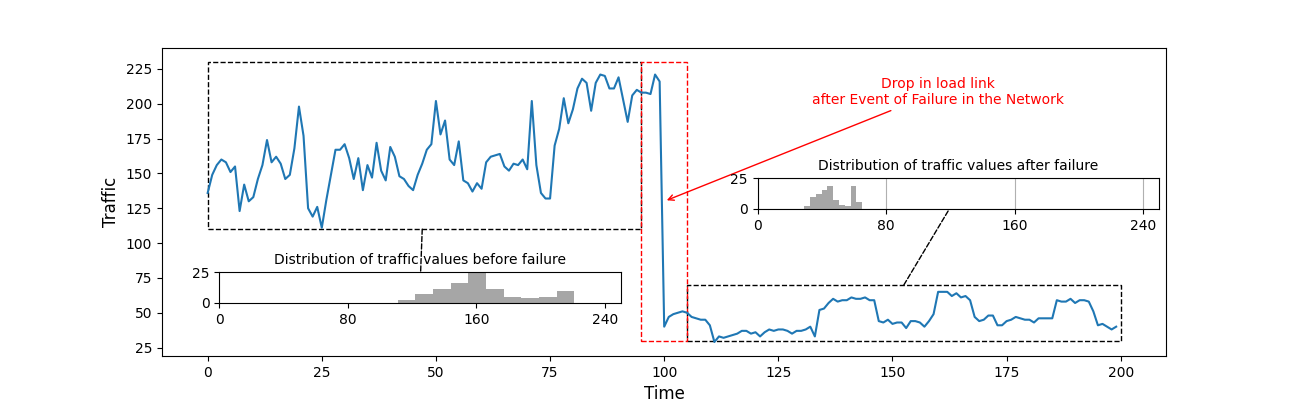}
  \caption{\small Example of a concept drift in traffic patterns and trends on a link due to network failure (at time step 100).}
  % \vspace{-0.8cm}
  \label{fig:illustrative}
\end{figure}

A potential solution to address this challenge is adopting a continual retraining process of \textsc{ml} models, as the most recent traffic measurements become available. While such an approach is effective under typical network (and thus, traffic) conditions, it fails to adapt in a timely manner under unforeseen network circumstances (and hence, under previously unseen network traffic conditions) for which the \textsc{ml} model was not originally trained. One such scenario arises during network faults, such as link failures, which can have cascading effects on traffic throughout the network, generating unprecedented patterns not encountered during the \textsc{ml} training process. In the context of \textsc{ml}, this shift in data patterns represent a phenomenon known as \emph{concept drift}, in which trained \textsc{ml} models fail when faced with new and unexpected data distributions \cite{lu2018learning}. Fig.~ \ref{fig:illustrative} shows the shifts in traffic patterns on a given link and the consequent distribution of traffic data (concept drift) resulting from a sudden decrease in traffic load due to a network failure (at time = 100) affecting another link. Note that an \textsc{ml} model developed to predict traffic is trained on data that shares a similar distribution to that seen prior to the failure.

The consequence of this shift in data distributions, or, in simpler terms, this \emph{mismatch}, between \textsc{ml} model training data and data seen in failure scenarios may cause, depending on the magnitude of this mismatch, a drastic deterioration in prediction accuracy, rendering the employed \textsc{ml} model ineffective. % or, in other words, out of service. 
The inability to accurately predict traffic under these circumstances poses a~severe obstacle to network operators, as timely and precise predictions are paramount for expediting the restoration of network elements and associated traffic flows. This issue underscores the critical need for a specialized focus on traffic prediction under unforeseen scenarios, such as in the case of a~failure of a~network element, where conventional models and approaches fall short. An advanced approach to tackle this issue is by relying on incremental learning strategies, which involves updating or expanding a model's knowledge over time without completely retraining the entire model. A primary limitation of this approach is its dependency on acquiring data that reflect the newly seen patterns for retraining. In a~scenario of network failure, an operator must first collect an adequate amount of data prior to retraining and redeploying the model. The temporal gap between the occurrence of the failure and the reintroduction of the freshly trained model imposes a~period of operational uncertainty, during which the operator cannot rely on an updated and reliable model. Note that frequent retraining, or retraining at shorter intervals, may not be a~feasible solution for operators to adequately address the task at hand, considering that the quantity (and quality) of data used may be insufficient. In fact, identifying the \emph{just-enough} quantity of data to collect (and hence, amount of waiting time) is crucial when adopting incremental learning approaches and merits considerable attention. However, even when this is determined, it does not eradicate the period of uncertainty that operators inevitably endure. Hence, employing methodologies that adapt seamlessly to shifts in data patterns under such failure scenarios, and consequently, allow operators to extract meaningful traffic predictions under a failure scenario until newly trained models are deployed, is essential.

In this work, we address the aforementioned challenge with a primary focus on the rapid adaptation of \textsc{ml} models for traffic prediction to drastic changes in traffic conditions arising from failure scenarios. Specifically, we propose a novel approach based on liquid neural networks (LNNs) \cite{hasani2021liquid}, which can adapt to changes in data patterns without need for retraining. We compare the performance of our proposed approach to a reference method based on incremental learning, which performs retraining periodically. To conduct our experiments, we propose and employ a traffic model and a restoration mechanism and simulate dynamic network operations in the event of a network failure impacting network traffic patterns. We present a comparative analysis of the two approaches in terms of their predictive performance and the time required to provide predictions %explanations 
within a predetermined error threshold. Experimental results show that LNN-based approaches provide great utility in scenarios characterized by abrupt shifts in traffic patterns whilst incremental learning-based approaches undergo retraining. The results also highlight the preference for extended intervals of periodic retraining of incremental learning approaches when faced with moderate changes in traffic patterns. 

The rest of the paper is organized as follows. Section \ref{related_work} discusses related work. Section \ref{methodology} formulates the problem of traffic prediction under failure scenarios and describes our proposed methodology to tackle it. Section \ref{modeling} describes traffic and network models adopted in our study. Section \ref{results} reports the experimental settings and discusses numerical results. Finally, Section \ref{conclusion} concludes the paper. 

\section{Background and Related Work}\label{related_work}
\subsection{Addressing Concept Drift}
Various techniques can be employed to address concept drift in \textsc{ml}. Retraining is a traditional technique that involves periodically updating \textsc{ml} models with the most recent data, allowing them to dynamically adjust to changing patterns. For instance, assuming that the data is a time series, \emph{windowing} or \emph{sliding window}, which is an approach that focuses on considering only a recent window of data for training, excluding older observations, can be employed. This technique enables models to swiftly adapt to the most recent patterns and minimizes the impact of outdated data, however, it still requires the availability of enough data for retraining. Another approach is \emph{incremental learning}, which supports updating models with new data without the need for retraining on the entire dataset, thus facilitating a more immediate response to changing patterns, yet still requiring the availability of new data. On the contrary, \emph{adaptive models}, such as online learning algorithms, are designed to inherently adjust their parameters as new data arrives, promoting continuous adaptation without the necessity for explicit retraining. In our work, we employ a novel online learning algorithm, i.e., LNNs, that does not require periodic updates to perform real-time adaptation. 

\subsection{ML-based Traffic Prediction}
Existing works on traffic prediction propose various statistical and \textsc{ml}-based methodologies for enhancing the predictive performance of their algorithms on short- and long-term traffic evolution \cite{lohrasbinasab2022statistical,ferreira2023forecasting, andreoletti2019network}. Since network traffic patterns and trends gradually change over extended periods, a specific focus should be given to dynamic approaches to predict previously unseen traffic conditions, as opposed to offline-learned models which fail to adapt in such cases \cite{knapinska2022long}. In particular, online traffic forecasting algorithms using data stream mining techniques were proposed as an effective solution to enable a gradual model adaptation over long periods \cite{shahraki2022comparative,liu2023multiclass,knapinska2022long,balanici2021classification}. Despite their effectiveness, these approaches are not seen fit to cope with rapid and drastic shifts in traffic patterns (i.e., with concept drift). To address this issue, incremental learning-based approaches have been recently proposed \cite{goscien2022efficient,knapinska2023link}, with a specific focus on forecasting traffic after a failure event in the network. More specifically, \cite{goscien2022efficient} proposes an algorithm based on \emph{moving windows}, demonstrating the trade-off between prediction quality and speed of adaptation. In \cite{knapinska2023link} authors employ \emph{partial fitting}, an incremental technique facilitating swift convergence after sudden traffic pattern changes due to network failure. This is achieved through model retraining with each batch of new data. While effective, the performance of this approach relies heavily on the batch size and retraining frequency, introducing uncertainty when significant changes in input data patterns occur in short periods. In this work, we tackle the problem from a different angle and propose a novel approach based on adaptive learning algorithm to adapt, in real time, to new data patterns without the need for retraining. We quantify the achievable advantages and identify the scenarios where such an approach can benefit the network operator with respect to the above-mentioned benchmark methods.

\section{Adaptive Learning and Incremental Learning for Link Load Prediction}\label{methodology}
The link load prediction task can be modeled as a regression problem which consists of forecasting the amount of traffic to be provisioned along a link in a future time step $t$, considering as input a set of $p$ observations of historical traffic measurements on the link. The success of the regressor (i.e., the link load predictor) is assessed by quantifying the deviation of traffic predictions with respect to the actual values. In the event of a failure in the network, which causes a shift in data distribution, the efficacy of the regression method is then quantified by its ability to promptly adapt to changes in traffic patterns (we introduce a metric explained in detail in Sec. \ref{settings}, referred to as \emph{Time to Convergence}, to quantify this adaptability). We consider two distinct approaches to address the problem at hand: $i$)  incremental learning techniques (reference scenarios) and $ii$) adaptive learning algorithms (proposed method). 

\subsection{Reference Approach: Incremental Learning}
The reference approach considered in this paper is based on the algorithm proposed in our previous work \cite{knapinska2023link}. To create a~model capable of adapting to changing traffic after failures, we developed an \emph{incremental learning} approach based on data stream mining techniques. To this end, we employ a~MultiLayer Perceptron (\textsc{mlp}) regressor that is periodically \emph{partially fitted}. In more detail, the model is first trained on a~number of traffic samples, and after enough new data arrives, it is updated to match the current traffic conditions. The size of the retraining window is a parameter that steers the frequency of the model partial fitting. 
As input features for the \textsc{mlp} learning algorithm we directly use raw data, i.e., the \emph{p} previous traffic samples. For example, model \emph{p3} implies that the prediction is made using three previous traffic samples as input features. 
Typically, streaming models utilize the \emph{test-then-train} protocol, i.e., for a specified batch size, the model outputs its forecast for the entire new batch of data, and when the real data is available – it undergoes retraining. However, such a~methodology does not allow for the direct use of previous samples as features, as they are not yet available to the model when making batch predictions. Therefore, the closest samples outside the prediction window act as the model inputs. In this work, we modify this approach to consent the best possible model adaptation after concept drifts. Specifically, the model predicts the traffic for the upcoming sample using \emph{p} previous ones. After the batch of new data is available, it undergoes retraining. 

\subsection{Proposed Approach: LNN-based Online Learning}

Artificial Neural Networks (\textsc{ann}s) are \textsc{ml} models inspired by the structure of the mammalian brain, which is composed of neurons organized in interconnected layers. In traditional \textsc{ann}s, neuron states are determined by linear combinations of inputs from other neurons, enhanced with specific non-linear activation functions (e.g., the sigmoid) to increase model expressiveness. A significant subtype of \textsc{ann}s, specifically designed to model time-series data, is the recurrent neural network (\textsc{rnn}). Such a network is characterized by recurrent mechanisms, that enable neuron activation to be influenced by linear combinations of inputs from other neurons and their own previous states. However, traditional \textsc{rnn}s often face challenges in adapting to complex time-series dynamics. \textsc{lnn}s \cite{hasani2021liquid} represent a fundamental shift from traditional \textsc{rnn}s. Indeed, while still making use of recurrent mechanisms, \textsc{lnn}s explicitly model time-series dynamics through differential equations that determine neuron states. Specifically, the neurons' state is the solution to the differential equation presented in Eq. \ref{eq:lnn}:

\begin{equation}\label{eq:lnn}
\frac{d\mathbf{x}(t)}{dt} = -\left[ \frac{1}{\tau} + f(\mathbf{x}(t), \mathbf{I}(t), \boldsymbol{\theta}) \right] \mathbf{x}(t) + f(\mathbf{x}(t), \mathbf{I}(t), \boldsymbol{\theta})\mathbf{A}
\end{equation}

where \( \mathbf{x}(t) \) is the vector representing the states of network neurons at time \( t \), \( \tau \) is a constant that ensures numerical stability, \( \mathbf{I}(t) \) represent the inputs to the neurons at time \( t \), \( f \) is a neural network parametrized by \( \boldsymbol{\theta} \) and \( \mathbf{A} \) is a bias term. This design offers more effective modeling of dynamic systems and possesses the remarkable capability to adapt \textsc{lnn}'s behavior post-training, i.e., they can adjust to the dynamics of unseen inputs without the need for further training. 

For this reason, \textsc{lnn}s are particularly useful in scenarios where sudden shifts in data distribution occur due to unforeseen events, such as a significant change in network traffic following a network failure. In such situations, network operators are required to respond quickly, basing their actions on the traffic load estimated by their models. However, models trained on prior data distributions may not provide reliable estimations after data distribution has changed. Typically, these models would necessitate re-training through incremental learning methods, which can only take place once an adequate volume of new data has been accumulated, thus further delaying network recovery. In contrast, models utilizing LNNs can adjust to novel data without the need for re-training, thereby facilitating a quicker response from network operators.

\section{Traffic and Network Model}\label{modeling}

\subsection{Traffic Model}\label{sec:traffic-model}
We use Euro28 topology (28 nodes, 82 directed links), which models the European core network~\cite{Orlowski_inoc07}. \textit{R} nodes selected based on real data\footnote{Available at http://www.datacentermap.com} host a~data center (\textsc{dc}) and provide anycast services/contents. We assume a~continuous inter-\textsc{dc}s synchronization, which guarantees that each \textsc{dc} offers exactly the same content and can serve any of the interested clients. However, we always assign a~client with the closest (according to the distance in kilometers) working \textsc{dc}. 

\begin{figure}
  \centering
  \includegraphics[height=6.5cm]{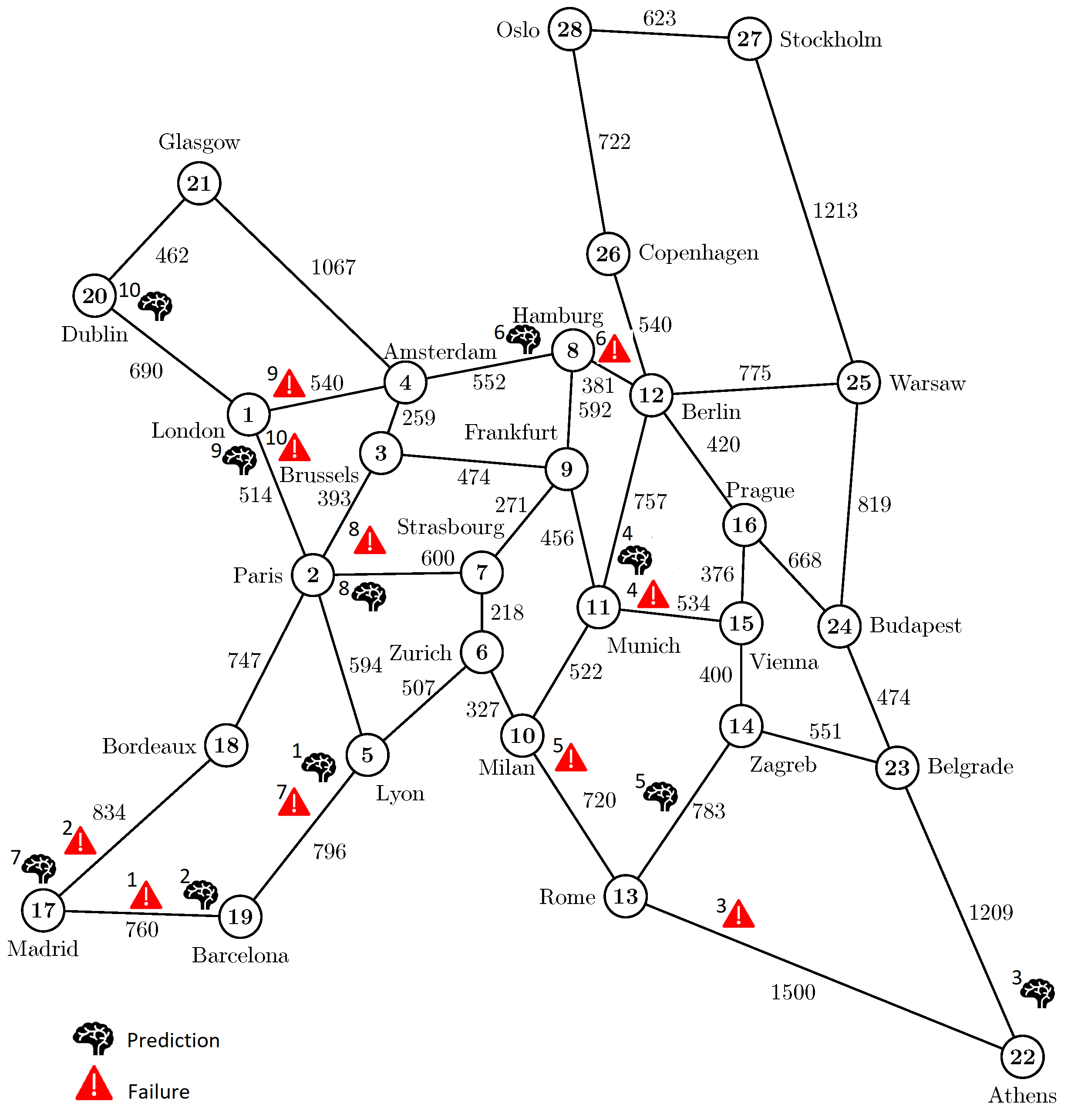}
  \caption{\small Topology of the network and failure scenarios.}
  \label{fig:euro28}
  % \vspace{-0.5cm}
\end{figure}

We work under the assumption that the network traffic is a~result of four transmission types, \textit{i})~city to city: a~basic unicast transmission between each pair of network nodes (representing cities), \textit{ii})~city to \textsc{dc}: a~service/content request and control data send from each city node to the closest working \textsc{dc}, \textit{iii}) \textsc{dc} to city: a~service/content provision and control data send from the selected \textsc{dc} to each city node, and \textit{iv}) \textsc{dc} to \textsc{dc}: a~unicast transmission observed between each pair of \textsc{dc}s; it realizes the inter-\textsc{dc}s synchronization. 

To mathematically describe the network traffic, we use the model proposed in \cite{Goscien_comnet23}. It models each transmission type (its time process) as a~sine (trigonometric) function with parameters (amplitude, pulsation, initial phase) determined by the network economical, demographic and topological parameters. The entire traffic between a~pair of nodes is then a~sum of the sine functions related to the transmission types observed between that pair. Considering a simulation consisting in \textit{T}-iterations (time steps), the model brings information about the traffic volume observed between each pair of network nodes for each time step $t \in T$. Within each time step of that time perspective, the average network load (the sum of all offered bitrate) is always equal to \textit{B} [Tbps].  

\subsection{Network Model}\label{sec:network-model}
Formally, a~network is modeled as a~directed graph $G=(V,E)$, where \emph{V} denotes a~set of nodes and \emph{E} indicates a~set of directed fiber links. Each fiber link offers the same spectrum range divided into \emph{S} frequency slices. Adjacent slices can be grouped then into spectral channels, each one characterized by the first slice index and the channel width. %\emph{R} network nodes, selected based on real data\footnote{Available at http://www.datacentermap.com} host a DC and can serve a~content/anycast service request. It is assumed that the content of all DCs is consistent over the entire network observation but each anycast client is always served by the closest working data center.

We consider a~dynamic network operation, which refers to the allocation of new arriving demands and resource release after expired demands at each time step. A~demand is given by a~tuple $d=(s,t,b,h)$, where $s(d)$ and $t(d)$ are demand's source and destination nodes, $b(d)$ is a~demand's volume (bitrate) in Gbps and $h(d)$ is a~demand holding time. Note that all demands are unicast, since the \textsc{dc} selection task is solved off-line and known in advance. The applied traffic model (see Sec~\ref{sec:traffic-model}) provides data regarding the total traffic bitrate observed between each pair of communicating nodes at each time step $t \in T$. To translate these series into sets of demands arriving at each time step, we make use a~special method introduced in~\cite{Goscien_rndm22}. At each time step $t \in T$, it iterates through all pairs of communicating nodes and divides the offered bitrate into a~set of arriving demands. To this end, it takes into account all previously offered and still existing demands. The method also assumes that the maximum demand's bitrate can be 250 Gbps while their duration is randomly selected from the range $(0,30]$ time steps. 

To deploy a~traffic demand, it is necessary to assign it with a~lightpath (a~routing path connecting the demand source and destination nodes and a~channel tailored to the demand's bitrate and path's length). A~demand can be allocated only at the time step of its arrival. If the available spectrum is not sufficient to serve it, the demand is rejected. 

%We assume that optical channels are multiplexed in a~flexible grid with a~standard slice width of 12.5~GHz. An~elastic transceiver operates at 28~Gbaud  with an optical channel bandwidth of 37.5~GHz (i.e., 3 frequency slices)~\cite{Ibrahimi_jocn21} and can use one of four modulation formats: \textsc{bpsk}, \textsc{qpsk}, 8-\textsc{qam}, 16-\textsc{qam}. 

We assume that optical channels are multiplexed in a~flexible grid with a~standard slice width of 12.5~GHz. An~elastic transceiver operates at 28~Gbaud  with an optical channel bandwidth of 37.5~GHz (i.e., 3 frequency slices) and can use one of four modulation formats: \textsc{bpsk}, \textsc{qpsk}, 8-\textsc{qam} and 16-\textsc{qam}. The {supported bitrates and transmission reaches} are {50 Gbps, 6300 km} for \textsc{bpsk}; {100 Gbps, 3500 km} for \textsc{qpsk}; {150 Gbps, 1200 km} for 8-\textsc{qam}; {200 Gbps, 600 km} for 16-\textsc{qam}.
%Table~\ref{tab:modulations} reports supported bit-rate and maximum transmission distance for each modulation~\cite{Khodashenas_lighttech16}. 
We allow for the usage of signal regenerators only if necessary, i.e., when a~path length exceeds the modulation reach. To select a~modulation and a~routing path for a~particular demand \textit{d}, we use the distance-adaptive transmission rule \cite{Walkowiak_springer16}. It chooses the most spectrally efficient format that simultaneously minimizes the number of used regenerators. %(\textsc{dat})

\begin{comment}
\begin{table}[htb]
    \centering
    \caption{Supported bitrate and transmission distance for a~transponder operating within 37.5~GHz spectrum \cite{Khodashenas_lighttech16}}
    \label{tab:modulations}
    \begin{footnotesize}
\begin{tabular}{c|cccc}
        \toprule
         & \textsc{bpsk} & \textsc{qpsk} &  \textsc{ 8-qam} & \textsc{ 16-qam} \\ \midrule
       supported bitrate [Gbps] & 50 & 100 & 150 & 200 \\ 
       transmission reach [km] & 6300 & 3500 & 1200 & 600 \\ 
        \bottomrule
    \end{tabular}
    \end{footnotesize}
\end{table} 
\end{comment}

The network operation is simulated within \textit{T} subsequent time steps. In each of them, a~set of demands arrives (and needs to be served) and a~subset of already allocated demands expires (and releases resources). When a~link failure occurs, the demands using that link (on their lightpaths) are affected and have to be restored. They are handled in the same way as new demands, however, their duration is shortened according to the time already spent in the network. The objective function is to serve as much bitrate as possible. However, the network load in this research is chosen to mitigate any demand blocking and to allow for full traffic restoration after a~link failure. 

\begin{figure*}[ht]
  \centering
  \begin{subfigure}{.24\linewidth}
    \resizebox{\linewidth}{!}{% This file was created with tikzplotlib v0.10.1.
\begin{tikzpicture}

\definecolor{darkgray176}{RGB}{176,176,176}
\definecolor{darkorange25512714}{RGB}{255,127,14}
\definecolor{forestgreen4416044}{RGB}{44,160,44}
\definecolor{lightgray204}{RGB}{204,204,204}
\definecolor{steelblue31119180}{RGB}{31,119,180}

\begin{axis}[
legend cell align={left},
legend style={fill opacity=0.8, draw opacity=1, text opacity=1, at={(0.9,0.9)}, draw=lightgray204},
tick align=outside,
tick pos=left,
x grid style={darkgray176},
xlabel={Timestep},
xmajorgrids,
xmin=971.5, xmax=1026.5,
xtick style={color=black},
xtick={974,984,994,1004,1014,1024},
xticklabels={0,10,20,30,40,50},
y grid style={darkgray176},
ylabel={RMSE},
ymajorgrids,
ymin=49.418594486682, ymax=185.762554817374,
ytick style={color=black}
]
\addplot [semithick, steelblue31119180]
table {%
974 174.927954637036
979 124.425580031766
984 103.460438398916
989 90.5538900529779
994 81.5999489030927
999 75.6030763290902
1004 70.4534795485521
1009 66.0498829329712
1014 62.3301218520857
1019 59.3791268304472
1024 57.00893997677
};
\addlegendentry{Incremental-5}
\addplot [semithick, darkorange25512714]
table {%
974 179.56510207507
979 127.873250072679
984 107.078532656948
989 93.6328679485466
994 84.5570932311824
999 78.3284242866765
1004 72.8710690072222
1009 68.2560555857113
1014 64.4277904726641
1019 61.4902733305455
1024 58.9442667112135
};
\addlegendentry{Incremental-20}
\addplot [semithick, forestgreen4416044]
table {%
974 168.982153273792
979 120.561125258283
984 101.235954995028
989 88.1456812311116
994 79.5673564888584
999 73.9281129469009
1004 68.610041588543
1009 64.2715848475339
1014 60.7064393458268
1019 58.047886980089
1024 55.6160472289862
};
\addlegendentry{LNN}
\end{axis}

\end{tikzpicture}}
    \caption{}
    \label{fig:sub1}
  \end{subfigure}%
  \hfill
  \begin{subfigure}{.24\linewidth}
    \resizebox{\linewidth}{!}{% This file was created with tikzplotlib v0.10.1.
\begin{tikzpicture}

\definecolor{darkgray176}{RGB}{176,176,176}
\definecolor{darkorange25512714}{RGB}{255,127,14}
\definecolor{forestgreen4416044}{RGB}{44,160,44}
\definecolor{lightgray204}{RGB}{204,204,204}
\definecolor{steelblue31119180}{RGB}{31,119,180}

\begin{axis}[
legend cell align={left},
legend style={fill opacity=0.8, draw opacity=1, text opacity=1, at={(0.9,0.9)}, draw=lightgray204},
tick align=outside,
tick pos=left,
x grid style={darkgray176},
xlabel={Timestep},
xmajorgrids,
xmin=971.5, xmax=1026.5,
xtick style={color=black},
xtick={974,984,994,1004,1014,1024},
xticklabels={0,10,20,30,40,50},
y grid style={darkgray176},
ylabel={MAPE},
ymajorgrids,
ymin=15.6272346648991, ymax=181.595982530989,
ytick style={color=black}
]
\addplot [semithick, steelblue31119180]
table {%
974 142.963008899395
979 76.4123098439359
984 58.602063319045
989 46.7218761742906
994 39.6798265389432
999 35.5622538473052
1004 32.0104487599217
1009 28.8282556398353
1014 26.1268010805554
1019 24.6754562138545
1024 23.1712686588123
};
\addlegendentry{Incremental-5}
\addplot [semithick, darkorange25512714]
table {%
974 174.051948537075
979 94.1906545004132
984 74.2387489511997
989 61.1837205543464
994 53.5378328836909
999 48.8722522977047
1004 44.0972699941303
1009 39.3024307130352
1014 35.5388268406794
1019 33.5805516919648
1024 31.7064162544416
};
\addlegendentry{Incremental-20}
\addplot [semithick, forestgreen4416044]
table {%
974 165.145598216312
979 89.2167130669213
984 69.0355868486094
989 54.4176277602362
994 46.3499291814623
999 42.0895361127149
1004 37.356151803193
1009 33.2005942187252
1014 30.0952901070314
1019 28.6874502769268
1024 26.9570128936152
};
\addlegendentry{LNN}
\end{axis}

\end{tikzpicture}}
    \caption{}
    \label{fig:sub2}
  \end{subfigure}%
  \hfill
  \begin{subfigure}{.24\linewidth}
    \resizebox{\linewidth}{!}{% This file was created with tikzplotlib v0.10.1.
\begin{tikzpicture}

\definecolor{darkgray176}{RGB}{176,176,176}
\definecolor{darkorange25512714}{RGB}{255,127,14}
\definecolor{forestgreen4416044}{RGB}{44,160,44}
\definecolor{lightgray204}{RGB}{204,204,204}
\definecolor{steelblue31119180}{RGB}{31,119,180}

\begin{axis}[
legend cell align={left},
legend style={fill opacity=0.8, draw opacity=1, text opacity=1, at={(0.9,0.9)}, draw=lightgray204},
tick align=outside,
tick pos=left,
x grid style={darkgray176},
xlabel={Timestep},
xmajorgrids,
xmin=971.5, xmax=1026.5,
xtick style={color=black},
xtick={974,984,994,1004,1014,1024},
xticklabels={0,10,20,30,40,50},
y grid style={darkgray176},
ylabel={RMSE},
ymajorgrids,
ymin=40.4211857208393, ymax=83.5737073492718,
ytick style={color=black}
]
\addplot [semithick, steelblue31119180]
table {%
974 81.553602158944
979 63.1484714203633
984 57.5486059472496
989 55.8934544332682
994 51.6257894018523
999 50.8810075229591
1004 49.7491195853137
1009 47.5101136720784
1014 46.4385526512224
1019 45.0677116370203
1024 43.6212732715719
};
\addlegendentry{Incremental-5}
\addplot [semithick, darkorange25512714]
table {%
974 79.5599828085986
979 61.5165354562707
984 55.4888785516795
989 54.0672888903873
994 49.4940612640568
999 49.3366813817497
1004 48.3192347528189
1009 46.26580621467
1014 45.0794226578922
1019 43.7279601344385
1024 42.3826639766771
};
\addlegendentry{Incremental-20}
\addplot [semithick, forestgreen4416044]
table {%
974 81.6122290934339
979 65.0586414375521
984 59.0880107112686
989 56.8582850336261
994 51.7864066587024
999 50.9159780565952
1004 49.6759345339943
1009 47.275007654717
1014 46.3184318431007
1019 44.7063036861122
1024 43.5104411083887
};
\addlegendentry{LNN}
\end{axis}

\end{tikzpicture}}
    \caption{}
    \label{fig:sub3}
  \end{subfigure}%
  \hfill
  \begin{subfigure}{.24\linewidth}
    \resizebox{\linewidth}{!}{% This file was created with tikzplotlib v0.10.1.
\begin{tikzpicture}

\definecolor{darkgray176}{RGB}{176,176,176}
\definecolor{darkorange25512714}{RGB}{255,127,14}
\definecolor{forestgreen4416044}{RGB}{44,160,44}
\definecolor{lightgray204}{RGB}{204,204,204}
\definecolor{steelblue31119180}{RGB}{31,119,180}

\begin{axis}[
legend cell align={left},
legend style={fill opacity=0.8, draw opacity=1, text opacity=1, at={(0.9,0.9)}, draw=lightgray204},
tick align=outside,
tick pos=left,
x grid style={darkgray176},
xlabel={Timestep},
xmajorgrids,
xmin=971.5, xmax=1026.5,
xtick style={color=black},
xtick={974,984,994,1004,1014,1024},
xticklabels={0,10,20,30,40,50},
y grid style={darkgray176},
ylabel={MAPE},
ymajorgrids,
ymin=8.82025591163754, ymax=16.8528717947018,
ytick style={color=black}
]
\addplot [semithick, steelblue31119180]
table {%
974 15.7203567421753
979 11.6072503585443
984 10.1897661665243
989 10.7892510201626
994 9.89391806170073
999 9.89051054910544
1004 10.2289517110888
1009 9.85726498695274
1014 9.79470082491837
1019 9.63897253019888
1024 9.35344567868378
};
\addlegendentry{Incremental-5}
\addplot [semithick, darkorange25512714]
table {%
974 15.0293610004731
979 11.0832055694311
984 9.627362453203
989 10.3940320303257
994 9.31714997647024
999 9.49753272068826
1004 9.89940210550445
1009 9.64943845064793
1014 9.5719465398539
1019 9.44326146204702
1024 9.18537481541319
};
\addlegendentry{Incremental-20}
\addplot [semithick, forestgreen4416044]
table {%
974 16.4877528909262
979 12.7195277661116
984 11.0511460940575
989 11.7662272441862
994 10.3657559508782
999 10.3837833514333
1004 10.6662049346218
1009 10.2492005573709
1014 10.1132984203592
1019 9.93330153081336
1024 9.68414938333938
};
\addlegendentry{LNN}
\end{axis}

\end{tikzpicture}}
    \caption{}
    \label{fig:sub4}
  \end{subfigure}
  \caption{\small The \textsc{rmse} and \textsc{mape} achieved by \emph{Online LNN}, \emph{Incremental-20} and \emph{Incremental-5} along the 50 time steps after the failure event (i.e., after concept drift) for the highly impacted (subfigures (a) and (b)) and moderately impacted (subfigures (c) and (d)) cases.}
  \label{fig:mainresults}
  % \vspace{-0.5cm}
\end{figure*}
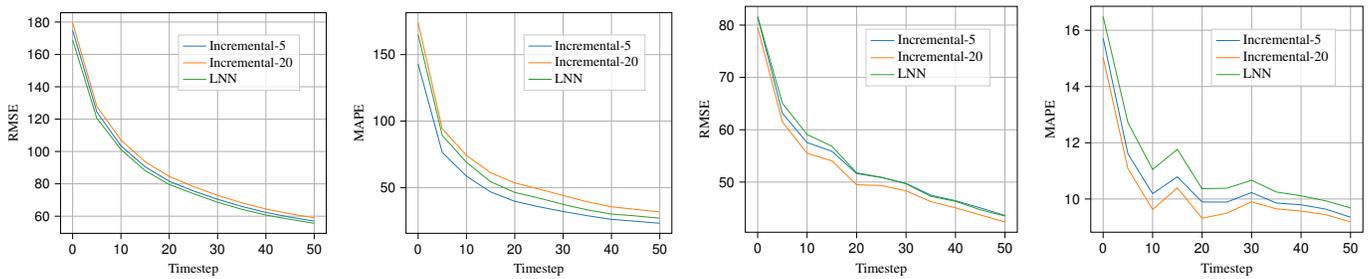

\section{Experimental Results}\label{results}
\subsection{Experimental Settings}\label{settings}
\subsubsection{Failure Scenarios}
We simulate 10 failure scenarios considering the topology depicted in Fig. \ref{fig:euro28} (see legend to identify failed link and inspected link pairs for each failure scenario). Upon a link failure event, the traffic restoration process takes place as described in Sec. \ref{sec:network-model}. Based on shift in data distribution of the traffic load on an inspected link, we divide the above 10 cases into two categories: 1) \emph{highly impacted} and 2) \emph{moderately impacted}. Specifically, highly impacted (resp., moderately impacted) category consists of cases in which the inspected link suffers from a drastic (moderate) variation of more (less) than 80\% between the mean value of the 50 observations just prior to failure and that of the 50 observations just after the failure. 

\subsubsection{Model Parameters and Training}
We consider a neural network composed of a first \textsc{lnn} layer with $30$ neurons and hyperbolic tangent activation function, followed by $3$ dense layers with a linear activation function and number of neurons of $10$, $5$ and $1$ (i.e., for the output layer), respectively.

We use the same \textsc{lnn} architecture for all links. %(for all cases). 
In all cases, we train the \textsc{lnn} using first 6000 time steps of the available data. The LNN is designed to take, at time step $t$, the past 3 observations from the previous three time steps ($t-3$, $t-2$ and $t-1$) to predict the traffic at time step $t+1$. Note that these 6000 observations (time steps) correspond to normal traffic conditions (prior to failure, with no concept drift). The \textsc{lnn} does not undergo any further training. We refer to this approach as $Online-{LNN}$. 

For the Incremental Learning approaches, we use an \textsc{mlp} with one hidden layer of twenty-five neurons, with the \textit{ReLU} activation function and \textit{adam} optimizer. Similarly to the case of \textsc{lnn}, we design the model to take as input, at time step $t$, the past 3 observations from previous three time steps ($t-3$, $t-2$ and $t-1$) to predict the traffic at time step $t+1$. We consider two variations of this model. The first undergoes batch retraining every 5 time steps while the second undergoes batch retraining every 20 time steps. We refer to them as $Incremental-5$ and $Incremental-20$, respectively.  

\subsubsection{Evaluation Metrics}
We consider three metrics to evaluate the performance of the proposed approaches, namely, the \emph{Root Mean Square Error} (\textsc{rmse}), the \emph{Mean Absolute Percentage Error} (\textsc{mape}) and the \emph{Time to Convergence} (TConv). 

The \textsc{rmse} evaluates how well the predicted values of the various approaches align with the actual observed traffic values in terms of their absolute values, penalizing larger errors more heavily, while the \textsc{mape} evaluates the percentage deviation between predicted and actual observed traffic values, to quantify the accuracy in a relative sense. In our work, we compute the \textsc{rmse} and the \textsc{mape} of the various approaches considering different time intervals, in terms of time steps, after the failure. This allows us to measure how the deviation between the predicted values of an approach and the actual values differ as we move farther from the failure. In other words, it allows us to identify when incremental learning methods, which undergo training, start to provide more accurate predictions than the LNN-based online learning approach. 

The \emph{TConv} is defined as the amount of time needed by an approach to provide \emph{x} number of predictions that are all within a predefined \emph{th} percentage error. We assume that \emph{th} represents a percentage error in the prediction that is tolerated by the operator, or in other words, a percentage error that minimally impacts the network operations that leverage the predictions. This allows us to identify when the incremental learning approaches provide reliable predictions and hence, the operator can again rely on them instead of LNN-based online learning approach. In our analysis we consider a fixed value of $x = 5$ and two distinct $th$: \(th \in \{10, 15\}\).

\subsection{Numerical Results and Discussion}
We start our discussion by comparing the performance of the three approaches in terms of \textsc{rmse} and \textsc{mape} considering the period after failure, in particular from the moment of failure (time step 0) to 50 time steps after failure (time step 50), as we aim to investigate how the various approaches react after the occurrence of concept drift. Figs.~\ref{fig:mainresults}(a) and \ref{fig:mainresults}(b) show the \textsc{rmse} and the \textsc{mape} of the three approaches in the highly impacted cases. In terms of \textsc{rmse}, \emph{\textsc{lnn}} outperforms the incremental-based approaches directly after failure (at time step 0), showing an \textsc{rmse} substantially lower (better) than \emph{Incremental 20} (168 instead of 180) and slightly lower than \emph{Incremental-5} (168 instead of 172). \emph{\textsc{lnn}} consistently outperforms incremental-based approaches in the subsequent 50 time steps, even though the incremental-based methods underwent multiple partial refitting processes up to that point. Specifically, \emph{Incremental-5} underwent 10 partial refitting processes, and \emph{Incremental-20} underwent two such processes. These results unveil two main findings in case of drastic change in traffic patterns: $i$) \emph{\textsc{lnn}} exhibits a consistent edge, albeit marginal, over incremental-based approaches. This advantage persists even when the incremental learning-based approaches undergo an intensive retraining process (every 5 time steps), for a substantial duration (e.g., up to 50 time steps), and $ii$) frequently performing partial refitting (considering a batch of 5) offers an improvement in the predictive quality of the models. 
In terms of \textsc{mape} (Fig. \ref{fig:mainresults}(b)), results show that \emph{Incremental-5} outperforms \emph{\textsc{lnn}} and \emph{Incremental-20}. While the results confirm the second finding observed when analyzing the performance in terms of \textsc{rmse}, they oppose the first one. This can be explained considering that the \textsc{mape} measures a relative error (i.e., to the actual traffic value), while the \textsc{rmse} measures an absolute error. Hence, large absolute errors may be absorbed by high reference traffic values, and are therefore more evident from the \textsc{rmse} than from the \textsc{mape}. We observe that, in the context of traffic allocation, absolute errors are more relevant than relative ones. For instance, the same \textsc{mape} value is more impactful on a~large traffic flow than on a small flow, as the actual traffic difference (e.g., the \textsc{rmse}) is much greater in the former case. In turn, underestimating or overestimating high traffic volumes (i.e., having a~large \textsc{rmse} but possibly a~small \textsc{mape}) could lead to significant problems such as congestion, under-utilization of resources, or even service outages. The choice of the metric to serve as foundation to select the approach to be adopted should be determined by the network managers, considering their specific objectives and network conditions.

Figures~\ref{fig:mainresults}(c) and \ref{fig:mainresults}(d) report the \textsc{rmse} and \textsc{mape} of the three approaches for the moderately impacted cases. Results show that, in terms of both metrics, \emph{Incremental-20} outperforms the \emph{\textsc{lnn}} and \emph{Incremental-5} directly after the failure (i.e., before \emph{Incremental-20} has undergone a partial refitting process) and throughout the considered period (up to time step 50). The \emph{\textsc{lnn}}, although very comparable to \emph{Incremental-5} in some cases, fails to edge any of the incremental learning-based approaches. We observe that in case of a relatively moderate change in data patterns, incremental learning approaches maintain their reliability over \emph{\textsc{lnn}}. This is attributed to their ability to leverage the knowledge accumulated through partial refitting, allowing them to adapt more effectively to the changing dynamics when the latter are not very drastic. It is important to note that in such scenarios, achieving better performance is observed when partial refitting is conducted less frequently, using a larger batch size of 20, as opposed to a more frequent approach with a smaller batch size of 5. This is expected, as larger batch sizes in partial refitting  allow the model to accumulate knowledge over a more extended period before updating, thus potentially capturing more stable patterns in the data and reducing overfitting to recent (and most likely, short-term) fluctuations (which are in fact very likely to occur just after the network has experienced a link failure). 

We now compare the approaches in terms of the \emph{TConv}. Table~\ref{tconv} reports the \emph{TConv} for each approach across the 10 cases, considering values of $x$ and $th$ as reported in Sec.~\ref{settings}. We first note how \emph{TConv} varies drastically among failure scenarios (e.g., for $th$ $=$ $10$, in failure scenario \#5 the various approaches require 32 time steps to converge while in failure scenario 6 they converge from the first time step). This highlights the complexity and variability of the problem. Comparing the various approaches, for $th = 10$, \textsc{lnn} shows the best \emph{TConv} in 6 out of 10 scenarios (in some scenarios exhibiting the same \emph{TConv} of incremental learning approaches). For $th = 15$, \textsc{lnn} shows the best \emph{TConv} in 50\% of the cases, achieving convergence in almost half the time with respect to the other approaches (e.g., failure scenario \#5). Despite \textsc{lnn}s efficacy in some scenarios, they struggle to outperform incremental learning in the rest of the scenarios. Therefore, a hybrid approach exploiting the strengths of each individual approaches could represent a promising solution. 

\begin{table}[]
\caption{\small TConv achieved by the various approaches across the 10 failure cases (see Fig. \ref{fig:euro28}) considering two distinct values of \emph{th}: 10 and 15, and for \emph{x} = 5.}
\scalebox{0.9}{%
\begin{tabular}{|l|l|llllllllll|}
\hline
\multicolumn{1}{|c|}{\multirow{2}{*}{th}} & \multicolumn{1}{c|}{\multirow{2}{*}{Approach}} & \multicolumn{10}{c|}{Failure Scenario ID}                                                                                                                                                                                                                                                                                                    \\ \cline{3-12} 
\multicolumn{1}{|c|}{}                    & \multicolumn{1}{c|}{}                          & \multicolumn{1}{c|}{1}          & \multicolumn{1}{c|}{2}           & \multicolumn{1}{c|}{3}          & \multicolumn{1}{c|}{4}           & \multicolumn{1}{c|}{5}           & \multicolumn{1}{c|}{6}          & \multicolumn{1}{c|}{7}          & \multicolumn{1}{c|}{8}          & \multicolumn{1}{c|}{9}          & \multicolumn{1}{c|}{10} \\ \hline
\multirow{3}{*}{\textit{10}}              & \textit{Incremental-5}                         & \multicolumn{1}{l|}{14}         & \multicolumn{1}{l|}{49}          & \multicolumn{1}{l|}{\textbf{2}} & \multicolumn{1}{l|}{\textbf{10}} & \multicolumn{1}{l|}{\textbf{32}} & \multicolumn{1}{l|}{\textbf{1}} & \multicolumn{1}{l|}{2}          & \multicolumn{1}{l|}{8}          & \multicolumn{1}{l|}{\textbf{4}} & \textbf{41}             \\ \cline{2-12} 
                                          & \textit{Incremental-20}                        & \multicolumn{1}{l|}{\textbf{6}} & \multicolumn{1}{l|}{\textbf{21}} & \multicolumn{1}{l|}{144}        & \multicolumn{1}{l|}{17}          & \multicolumn{1}{l|}{\textbf{32}} & \multicolumn{1}{l|}{\textbf{1}} & \multicolumn{1}{l|}{17}         & \multicolumn{1}{l|}{8}          & \multicolumn{1}{l|}{5}          & \textbf{41}             \\ \cline{2-12} 
                                          & \textit{LNN}                                   & \multicolumn{1}{l|}{9}          & \multicolumn{1}{l|}{\textbf{21}} & \multicolumn{1}{l|}{35}         & \multicolumn{1}{l|}{18}          & \multicolumn{1}{l|}{\textbf{32}}          & \multicolumn{1}{l|}{\textbf{1}} & \multicolumn{1}{l|}{\textbf{1}} & \multicolumn{1}{l|}{\textbf{5}} & \multicolumn{1}{l|}{\textbf{4}} & 74                      \\ \hline
\multirow{3}{*}{\textit{15}}              & \textit{Incremental-5}                         & \multicolumn{1}{l|}{7}          & \multicolumn{1}{l|}{\textbf{2}}  & \multicolumn{1}{l|}{\textbf{2}} & \multicolumn{1}{l|}{2}           & \multicolumn{1}{l|}{32}          & \multicolumn{1}{l|}{\textbf{1}} & \multicolumn{1}{l|}{\textbf{1}} & \multicolumn{1}{l|}{\textbf{4}} & \multicolumn{1}{l|}{3}          & 10                      \\ \cline{2-12} 
                                          & \textit{Incremental-20}                        & \multicolumn{1}{l|}{\textbf{6}} & \multicolumn{1}{l|}{7}           & \multicolumn{1}{l|}{34}         & \multicolumn{1}{l|}{2}           & \multicolumn{1}{l|}{31}          & \multicolumn{1}{l|}{\textbf{1}} & \multicolumn{1}{l|}{\textbf{1}} & \multicolumn{1}{l|}{5}          & \multicolumn{1}{l|}{\textbf{1}} & \textbf{9}              \\ \cline{2-12} 
                                          & \textit{LNN}                                   & \multicolumn{1}{l|}{\textbf{6}} & \multicolumn{1}{l|}{7}           & \multicolumn{1}{l|}{14}         & \multicolumn{1}{l|}{\textbf{0}}  & \multicolumn{1}{l|}{\textbf{17}} & \multicolumn{1}{l|}{\textbf{1}} & \multicolumn{1}{l|}{\textbf{1}} & \multicolumn{1}{l|}{5}          & \multicolumn{1}{l|}{3}          & 41                      \\ \hline
\end{tabular}
}
\label{tconv}
\end{table}

\section{Conclusion}\label{conclusion}
In this paper, we addressed the concept drift issue in traffic patterns arising due to network failure for machine learning-based link load prediction. To this end, we propose an adaptive learning approach based on the use of liquid neural networks to adapt to changes in traffic patterns without requiring any retraining. As reference scenario, we design incremental learning-based approaches that undergo partial refitting periodically. We compare our \textsc{lnn}-based proposed approach to incremental learning-based approaches in terms of quality of their predictions and time to adapt to newly-seen traffic patterns. Our results show that \textsc{lnn}-based approaches can come in handy in circumstances of drastic change in traffic patterns, whilst incremental learning-based approaches can be retrained and adapted. Results also reveal that a larger interval for periodic refitting is desirable when change in traffic patterns is relatively moderate. Overall, our results provide network managers with valuable insights for machine learning-based traffic prediction in case of network failure.
\bibliographystyle{IEEEtran}
\bibliography{bibliography}
\end{document}